\documentclass[aip,jcp,reprint,noshowkeys,superscriptaddress]{revtex4-1}
\usepackage{graphicx,dcolumn,bm,xcolor,microtype,multirow,amscd,amsmath,amssymb,amsfonts,physics,longtable,wrapfig,bbold,siunitx,xspace}
\usepackage[version=4]{mhchem}

\usepackage[utf8]{inputenc}
\usepackage[T1]{fontenc}

\usepackage{hyperref}
\hypersetup{
    colorlinks,
    linkcolor={red!50!black},
    citecolor={red!70!black},
    urlcolor={red!80!black}
}

\usepackage{listings}
\definecolor{codegreen}{rgb}{0.58,0.4,0.2}
\definecolor{codegray}{rgb}{0.5,0.5,0.5}
\definecolor{codepurple}{rgb}{0.25,0.35,0.55}
\definecolor{codeblue}{rgb}{0.30,0.60,0.8}
\definecolor{backcolour}{rgb}{0.98,0.98,0.98}
\definecolor{mygray}{rgb}{0.5,0.5,0.5}

\definecolor{sqred}{rgb}{0.85,0.1,0.1}
\definecolor{sqgreen}{rgb}{0.25,0.65,0.15}
\definecolor{sqorange}{rgb}{0.90,0.50,0.15}
\definecolor{sqblue}{rgb}{0.10,0.3,0.60}

\lstdefinestyle{mystyle}{
    backgroundcolor=\color{backcolour},
    commentstyle=\color{codegreen},
    keywordstyle=\color{codeblue},
    numberstyle=\tiny\color{codegray},
    stringstyle=\color{codepurple},
    basicstyle=\ttfamily\footnotesize,
    breakatwhitespace=false,
    breaklines=true,
    captionpos=b,
    keepspaces=true,
    numbers=left,
    numbersep=5pt,
    numberstyle=\ttfamily\tiny\color{mygray},
    showspaces=false,
    showstringspaces=false,
    showtabs=false,
    tabsize=2
  }

  \usepackage[]{notes2bib}

  \newcolumntype{d}{D{.}{.}{-1}}

\newcommand{\alert}[1]{\textcolor{black}{#1}}

\newcommand{\ie}{\textit{i.e.}\xspace}

\newcommand{\hH}{\Hat{H}} 
\newcommand{\hT}{\Hat{T}} 
\newcommand{\hV}{\Hat{V}} 
\newcommand{\hn}{\Hat{n}}
\newcommand{\cre}[1]{a_{#1}^\dagger}
\newcommand{\ani}[1]{a_{#1}} 

\newcommand{\br}{\boldsymbol{r}} 

\newcommand{\up}{\uparrow} 
\newcommand{\dw}{\downarrow} 
\newcommand{\Dv}{\Delta v} 
\newcommand{\Dn}{\Delta n} 
\newcommand{\stat}[1]{\underset{#1}{\text{stat}}}

\newcommand{\LCPQ}{Laboratoire de Chimie et Physique Quantiques (UMR 5626), Universit\'e de Toulouse, CNRS, UPS, France}

\begin{document}	

\title{Exact Excited-State Functionals of the Asymmetric Hubbard Dimer}

\author{Sara \surname{Giarrusso}}
	\email{sgiarrusso@irsamc.ups-tlse.fr}
	\affiliation{\LCPQ}

\author{Pierre-Fran\c{c}ois \surname{Loos}}
	\email{loos@irsamc.ups-tlse.fr}
	\affiliation{\LCPQ}

\begin{abstract}
\textbf{Abstract:} The exact functionals associated with the (singlet) ground and the two singlet excited states of the asymmetric Hubbard dimer at half-filling are calculated using both Levy's constrained search and Lieb's convex formulation.
While the ground-state functional is, as commonly known, a convex function with respect to the density, the functional associated with the doubly-excited state is found to be concave.
Also, because the density-potential mapping associated with the first excited state is non-invertible, its ``functional'' is a partial, multi-valued function composed of one concave and one convex branch that correspond to two separate domains of values of the external potential.
Remarkably, it is found that, although the one-to-one mapping between density and external potential may not apply (as in the case of the first excited state), each state-specific energy and corresponding universal functional are ``functions'' whose derivatives are each other's inverse, just as in the ground state formalism.
These findings offer insight into the challenges of developing state-specific excited-state density functionals for general applications in electronic structure theory.
\bigskip
\begin{center}
	\boxed{\includegraphics[width=0.5\linewidth]{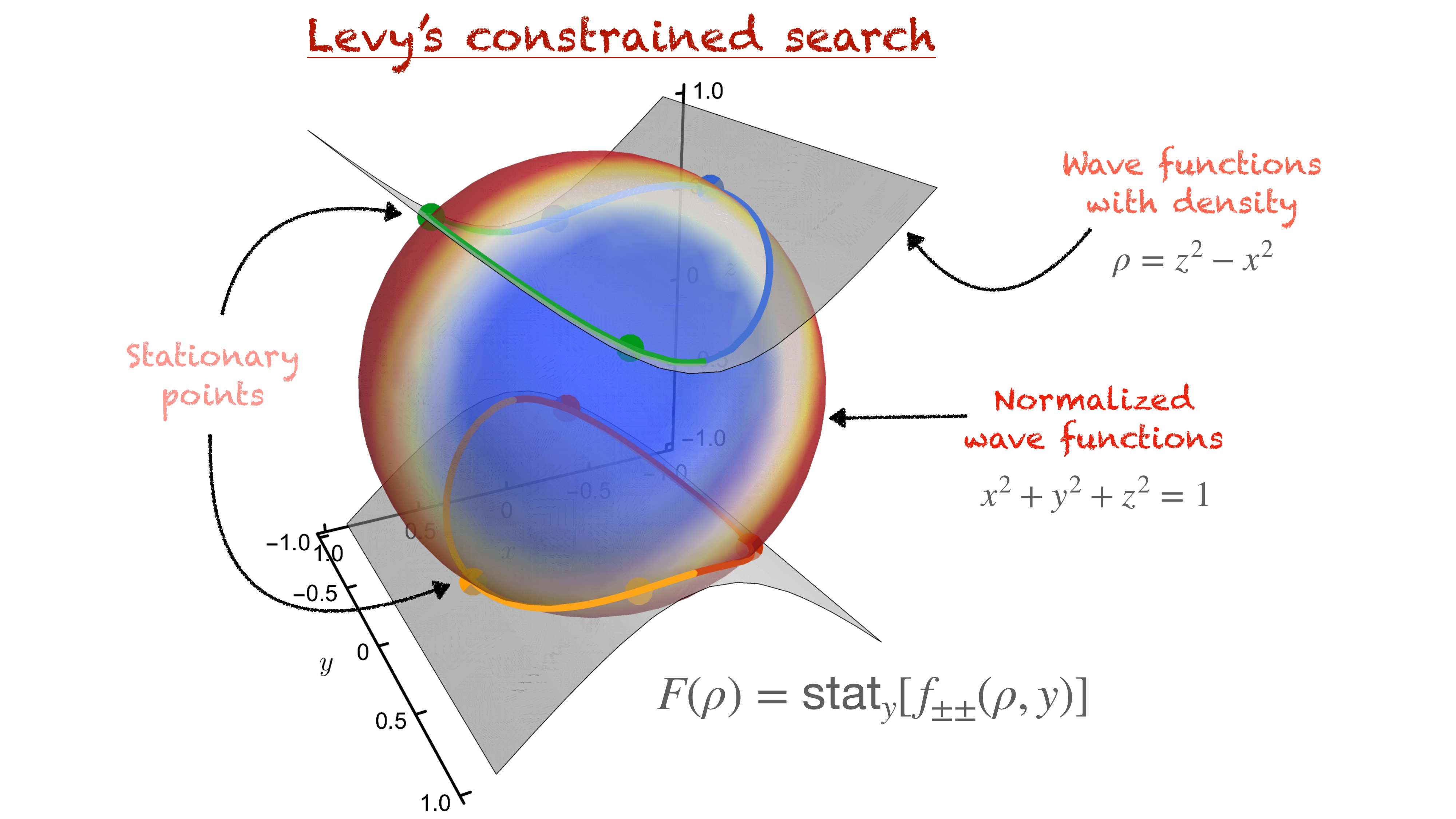}}
\end{center}
\bigskip
\end{abstract}

\maketitle

Several decades after its foundation, \cite{Hohenberg_1964} density-functional theory (DFT) still represents the main computational tool to perform quantum mechanical simulations of interest for pharmaceutical and technological applications. \cite{Teale_2022} Originally developed as a ground-state theory, it has been swiftly extended to calculate the lowest excited state of a given symmetry, \cite{Gunnarsson_1976,Ziegler_1977,Gunnarsson_1979,vonBarth_1979,Englisch_1988} thereby obtaining excitation energies from differences of self-consistent field ($\Delta$SCF) calculations.

Notwithstanding the usefulness of such extension, for more general purposes, one usually relies on (linear-response) time-dependent (TD) DFT to describe excited states at the DFT level. \cite{Runge_1984,Appel_2003, Burke_2005,Casida_2012,Huix-Rotllant_2020} TDDFT is an in-principle exact theory but, in practice, it relies on approximations for the exchange-correlation kernel. A fundamental source of error underlying virtually all its implementations is adiabaticity (neglecting memory effects), while another type of error comes from the particular choice of the exchange-correlation functional, similar to ground-state Kohn-Sham DFT. \cite{Kohn_1965} Within these approximations, TDDFT has seen important successes \cite{Jacquemin_2009} but is also plagued by well-known shortcomings, e.g., for the description of double excitations or charge-transfer processes. \cite{Tozer_1999,Tozer_2000,Dreuw_2003,Maitra_2004,Levine_2006,Maitra_2017}

Due to the relevance of these phenomena in photochemical applications or quantum-based technologies, alternative, time-independent theories have been developed. The most well-known is ensemble DFT (EDFT), based on an ensemble of equally-weighted \cite{Theophilou_1979} or unequally-weighted \cite{Gross_1988a,Gross_1988b,Oliveira_1988} densities, each coming from an individual quantum state rather than a pure-state density as in traditional DFT. In recent times, EDFT and related theories have undergone significant developments that are crucial to its advancement. \cite{Pribram-Jones_2014,Yang_2014,Yang_2017,Sagredo_2018,Filatov_2016,Senjean_2015,Deur_2017,Deur_2018,Deur_2019,Marut_2020,Loos_2020g,Fromager_2020,Cernatic_2022,Gould_2017,Gould_2018,Gould_2019,Gould_2020,Gould_2021,Gould_2022,Gould_2023,Schilling_2021,Liebert_2021,Liebert_2022,Liebert_2023a,Liebert_2023b} However, it suffers from the disadvantages that, to treat a high-lying excited state, \alert{it is usually required to include all lower-lying states in the ensemble}, and that the weight dependence of the exchange-correlation functional is hard to model. Another ensemble theory that has been receiving increasing attention and shares some of the problems of EDFT is $w$-ensemble one-body reduced density matrix functional theory. \cite{Schilling_2021,Liebert_2021,Liebert_2022,Liebert_2023a,Liebert_2023b}

Concerning pure excited states, orbital-optimized DFT, \cite{Perdew_1985,Kowalczyk_2013,Gilbert_2008,Barca_2018a,Barca_2018b,Hait_2020,Hait_2021,Shea_2018,Shea_2020,Hardikar_2020,Levi_2020,Carter-Fenk_2020,Toffoli_2022,Schmerwitz_2023} the extension to any excited state of the mentioned $\Delta$SCF calculations, has been shown to be relatively successful for the calculation of classes of excitations where TDDFT typically fails, \cite{Hait_2020,Hait_2021} although its theoretical underpinning is still in progress.

From a theoretical perspective, state-specific density-functional formalisms have been developed. \cite{Gorling_1996,Nagy_1998,Levy_1999,Gorling_1999,Zhang_2004,Ayers_2009,Ayers_2012,Ayers_2015,Ayers_2018,Garrigue_2022}
Some of these are complicated by the dependence of the functional on quantities other than the excited-state density and/or by the need for orthogonality constraints to inherit the variational character of the ground-state theory. \cite{Lieb_1985}

In his seminal work, G\"orling \cite{Gorling_1999} proposes a \textit{stationarity} rather than a \textit{minimum} principle to treat excited states. Building on G\"orling's work \cite{Ayers_2009} and restricting the set of external potentials to Coulombic ones, Ayers \textit{et al.} establish a one-to-one mapping between external potential and any of its associated stationary densities. \cite{Ayers_2012,Ayers_2015,Ayers_2018} For a general external potential, this one-to-one mapping may not hold true. \cite{Perdew_1985,Gaudoin_2004,Samal_2005,Samal_2006}
However, none of these formalisms have revealed a fundamental dual relationship between excited-state energy and its corresponding state-specific functional similar to the one between the ground-state energy and the universal functional elucidated by Lieb. \cite{Lieb_1983}

In turn, the present Letter provides an explicit case in which such a fundamental dual relationship carries through for excited states. Adopting G\" orling's stationarity principle \cite{Gorling_1999} on Levy's constrained search \cite{Levy_1979} and Lieb's convex formulation, \cite{Lieb_1983} we find for a simple model that, just as for the ground state, a given excited-state energy and its corresponding universal functional are functions whose derivatives are each other's inverse functions, a property described as ``the essence of DFT''. \cite{Helgaker_2022} Yet the ``functional'' associated with the first-excited state has some very peculiar mathematical properties.

Below, we first review the ground-state formalism. Consider the usual variational principle
\begin{equation}\label{eq:RaleighRitz}
	E[v] = \min_{\Psi} \mel{\Psi}{\hH_v}{\Psi}
\end{equation}
where the minimization is performed over all normalized $N$-electron antisymetrized wave functions $\Psi$ and the electronic Hamiltonian
\begin{equation}
	\hH_v = \hT + \hV_{ee} + \sum_{i=1}^N v(\br_i)
\end{equation}
is composed of the kinetic energy operator $\hT$, the electron repulsion operator $\hV_{ee}$, and the external potential contribution.

The minimization in Eq.~\eqref{eq:RaleighRitz} can be split in two steps
\begin{equation}
\label{eq:HKvar}
\begin{split}
	E[v] & = \min_{\rho} \min_{\Psi \leadsto\rho} \mel{\Psi}{\hH_v}{\Psi} 
	\\
	& = \min_{\rho} \qty{ F[\rho] + \int v(\br) \rho(\br) d\br }
\end{split}
\end{equation}
where in the second line we have introduced the Levy-Lieb or ``universal'' functional defined, via Levy's constrained search,  \cite{Levy_1979} as
\begin{equation}
	F[\rho] 
	= \min_{\Psi \leadsto \rho} \mel{\Psi}{\hH_0}{\Psi}
	= \mel{\Psi[\rho]}{\hH_0}{\Psi[\rho]}
\end{equation}
Note that the Hohenberg-Kohn, \cite{Hohenberg_1964} Levy-Lieb, \cite{Levy_1979,Lieb_1983} or Lieb functional \cite{Lieb_1983} differ in the density domain. We refer to any of them as the universal functional, although only the Lieb functional is properly convex in $\rho$. \cite{Helgaker_2022}

The Legendre-Fenchel transform of Eq.~\eqref{eq:HKvar} delivers $F[\rho]$ from the maximisation
\begin{equation}
	\label{eq:Lvar}
	F[\rho] = \max_{v}\qty{ E[v] - \int v(\br) \rho(\br) d\br }
\end{equation}
exemplifying the duality between the functional $E[v]$, concave in the external potential $v$, and $F[\rho]$, convex in the density $\rho$. \cite{Lieb_1983}
Although technically discontinuous, $F[\rho]$ is ``almost differentiable'' \cite{Helgaker_2022} in that it may be approximated to any accuracy by a differentiable regularized functional. \cite{Kvaal_2014} Thus, assuming differentiability and carrying out the optimizations in Eqs.~\eqref{eq:HKvar} and \eqref{eq:Lvar}, one obtains
\begin{subequations}
\begin{align}
	&\fdv{F[\rho(\br)]}{\rho(\br)} + v(\br) = 0 
	\label{eq:EulerDFT}
	\\
	&\fdv{E[v(\br)]}{v(\br)} - \rho(\br) = 0 
	\label{eq:EulerPFT}
\end{align}
\end{subequations}
respectively.

\begin{figure}
	\includegraphics[width=\linewidth]{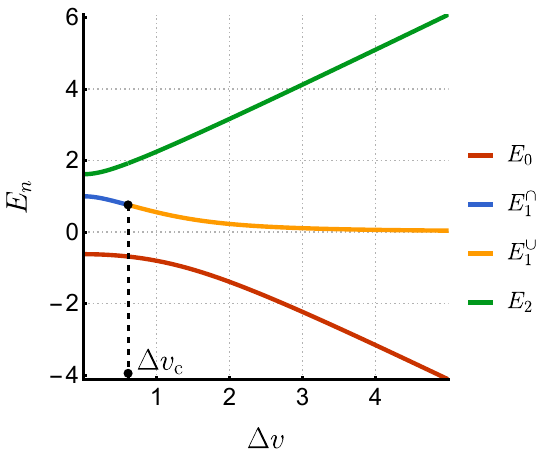}
	\caption{$E_0$, $E_1$, and $E_2$ as functions of $\Dv$ for $t = 1/2$ and $U = 1$.
	Note that $E$ is an even function of $\Dv$.
	$E_1$ is concave for $\Dv < \Dv_\text{c}$ and becomes convex for larger $\Dv$ values.}
	\label{fig:energies}
\end{figure}

\begin{figure}
	\includegraphics[width=\linewidth]{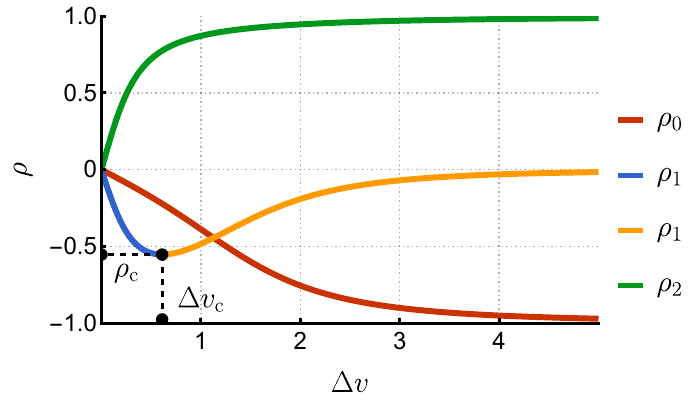}
	\caption{$\rho$ as a function of $\Dv$ for $t = 1/2$ and $U = 1$ for the ground-state ($\rho_0$), the singly-excited state ($\rho_1$), and the doubly-excited states ($\rho_2$).
	$\rho_1$ reaches a critical value, $\rho_\text{c}$, at $\Dv_\text{c}$.	
	Note that $\rho$ is an odd function of $\Dv$.}
	\label{fig:densities}
\end{figure}

We adopt the two-site Hubbard model at half-filling, \cite{Hubbard_1963,Lieb_1968,Schonhammer_1987,Montorsi_1992,Carrascal_2015,Cohen_2016,Ying_2016,Smith_2016,Senjean_2017,Deur_2017,Carrascal_2018} whose Hamiltonian reads 
\begin{equation}
    \hH = 
    - t \sum_{\sigma=\up,\dw} \qty( \cre{0\sigma} \ani{1\sigma} + \text{h.c.} )
    + U \sum_{i=0}^{1} \hat{n}_{i\up} \hat{n}_{i\dw} 
	+  \Dv \frac{\hn_{1} - \hn_{0}}{2}
\end{equation}
where $t > 0$ is the hopping parameter, $U \ge 0$ is the on-site interaction parameter, $\hn_{i\sigma} = \cre{i\sigma}\ani{i\sigma}$ is the spin density operator on site $i$, $\hn_{i} = \hn_{i\uparrow} + \hn_{i\downarrow}$ is the density operator on site $i$, and $\Dv = v_{1}-v_{0}$ (with $v_0 + v_1 = 0$) is the potential difference between the two sites. 

Although simple, this model is able to describe the physics of partially-filled narrow band gaps \cite{Hubbard_1963,Lieb_1968,Montorsi_1992} and its two-site version has been used in the framework of site-occupation function theory to exemplify central concepts or test (new) density-functional methods by numerous authors. \cite{Schonhammer_1987,Carrascal_2015,Cohen_2016,Ying_2016,Smith_2016,Senjean_2017,Deur_2017,Carrascal_2018}

\alert{It is noteworthy to mention that for lattice systems, even in the case of the ground-state functional, the Hohenberg-Kohn theorem does not hold universally. While a potential does exist, it is not always unique. This aspect was recently highlighted by Penz and van Leeuwen. \cite{Penz_2021} However, in the context of linear Hubbard chains (like the one discussed in this paper where the chain is of length two), the uniqueness and thus the applicability of the Hohenberg-Kohn theorem is established by Theorem 17 in the aforementioned reference. This theorem provides a robust guarantee of uniqueness, ensuring that for linear Hubbard chains, there is a unique potential associated with a given density. Notably, the present study and Sch\"onhammer and Gunnarsson's original work also support this. \cite{Schonhammer_1987}}

At half filling ($N=2$), we expand the Hamiltonian in the $N$-electron (spin-adapted) site basis $\ket{0_\up0_\dw}$, $(\ket{0_\up1_\dw}- \ket{0_\dw1_\up})/\sqrt{2}$, and $\ket{1\up1\dw}$ to form the following Hamiltonian matrix
\begin{equation}
	\vb{H} = \mqty(
		U - \Dv & -\sqrt{2} t  & 0 \\
		-\sqrt{2} t & 0 & -\sqrt{2}t \\
		0 & -\sqrt{2}t & U+\Dv \\
		) 
\end{equation}
whose eigenvalues provide the singlet energies of the system.
A generic singlet wave function can then be written as 
\begin{equation}
	\ket{\Psi} = x \ket{0_\up0_\dw} + y \frac{\ket{0_\up1_\dw} - \ket{0_\dw1_\up}}{\sqrt{2}} + z \ket{1_\up1_\dw}
\end{equation}
with $-1 \le x,y,z \le 1$ and the normalization condition 
\begin{equation}
\label{eq:normalization}
	x^2 + y^2 + z^2 = 1
\end{equation}
The energy is given by $E = T + V_{ee} + V$, with
\begin{subequations}
\begin{align}
	\label{eq:T}
	&T  = - 2\sqrt{2} t y \qty(x + z)
	\\
	\label{eq:Vee}
	&V_{ee}  = U \qty(x^2 + z^2)
	\\
	\label{eq:V}
	&V  = \rho \, \Dv  
\end{align}
\end{subequations}
with
\begin{equation}
	\rho = \mel{\Psi}{\frac{\hn_{1} - \hn_{0}}{2}}{\Psi} =  (z^2 - x^2)
\end{equation}

We call $E_0$, $E_1$, and $E_2$ the energies of the ground state, first (singly-)excited state, and second (doubly-)excited state, respectively. These are represented in Fig.~\ref{fig:energies} as functions of $\Dv$ for $t=1/2$ and $U=1$. It is worth noting that $E_0$ (red curve) and $E_2$ (green curve) are concave and convex with respect to $\Dv$, respectively, for any value of $t$ and $U$, while $E_1$ is concave for $\Dv$ smaller than a critical value $\Dv_\text{c}$ (blue curve labeled as $E_1^{\cap}$) and becomes convex for $\Dv > \Dv_\text{c}$ (yellow curve labeled as $E_1^{\cup}$).

The corresponding differences in (reduced) site occupation
\begin{equation}
\label{eq:rho}
	\rho = \frac{\Dn}{2}
\end{equation}
for the ground state, $\rho_0$, first excited state, $\rho_1$, and second excited state, $\rho_2$, are represented in Fig.~\ref{fig:densities}. While the ground (red curve) and the doubly-excited (green curve) states have monotonic densities with respect to $\Dv$ for any $t$ and $U$ values, $\rho_1$ is non-monotonic and reaches a critical value $\rho_\text{c}$ at $\Dv_\text{c}$ before decaying to $0$ as $\Dv\to\infty$. In agreement with Eq.~\eqref{eq:EulerPFT}, in the asymmetric Hubbard dimer, one finds
\begin{equation}
         \dv{E_0(\Dv)}{\Dv} = 2\, \rho_0(\Dv)
         \end{equation}
However, analogous relations hold true also for the two excited states, i.e.,
\begin{subequations}
\begin{align}
	\dv{E_1(\Dv)}{\Dv} & = 2\, \rho_1(\Dv)\\
	\dv{E_2(\Dv)}{\Dv} & = 2\, \rho_2(\Dv)
\end{align}
\end{subequations}
\\
\begin{figure*}
	\includegraphics[width=\linewidth]{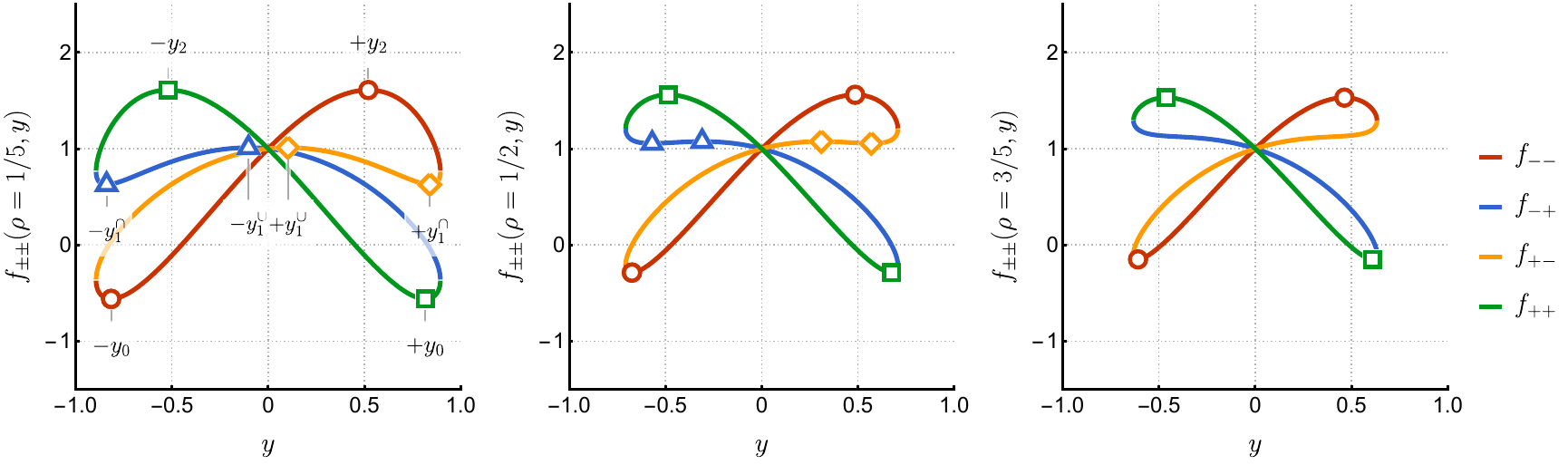}
	\caption{$f_{--}(\rho,y)$ (red), $f_{-+}(\rho,y)$ (blue), $f_{+-}(\rho,y)$ (yellow), and $f_{++}(\rho,y)$ (green) as functions of $y$ for $t = 1/2$, $U = 1$, and $\rho = 1/5$ (left), $1/2$ (center), and $3/5$ (right).
	The markers indicate the position of the stationary points on each branch.
	At $\rho = 3/5$ (right panel), the stationary points of $f_{-+}$ and $f_{+-}$ have disappeared as $\rho > \rho_c$ (see Fig.~\ref{fig:densities}).}
	\label{fig:fpm}
\end{figure*}

\begin{figure}
	\includegraphics[width=\linewidth]{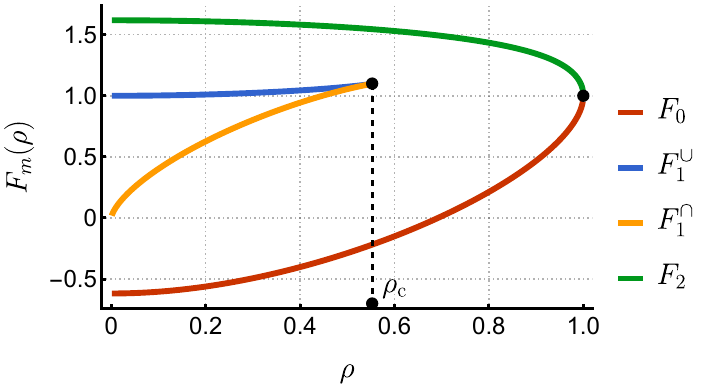}
	\caption{State-specific exact functionals $F_m(\rho)$ as functions of $\rho$ for $t = 1/2$ and $U = 1$.
	The ground-state functional $F_0(\rho)$ (red) is \alert{convex} with respect to $\rho$, the singly-excited state multi-valued functional $F_1(\rho)$ has one \alert{convex} branch (blue) and one \alert{concave} branch (yellow), each associated with a separate set of $\Dv$ values, while the doubly-excited state functional $F_2(\rho)$ (green) is \alert{concave}.
	Note that $F$ is an even function of $\rho$.
	}
	\label{fig:F}
\end{figure}

\begin{figure}
	\includegraphics[width=0.7\linewidth]{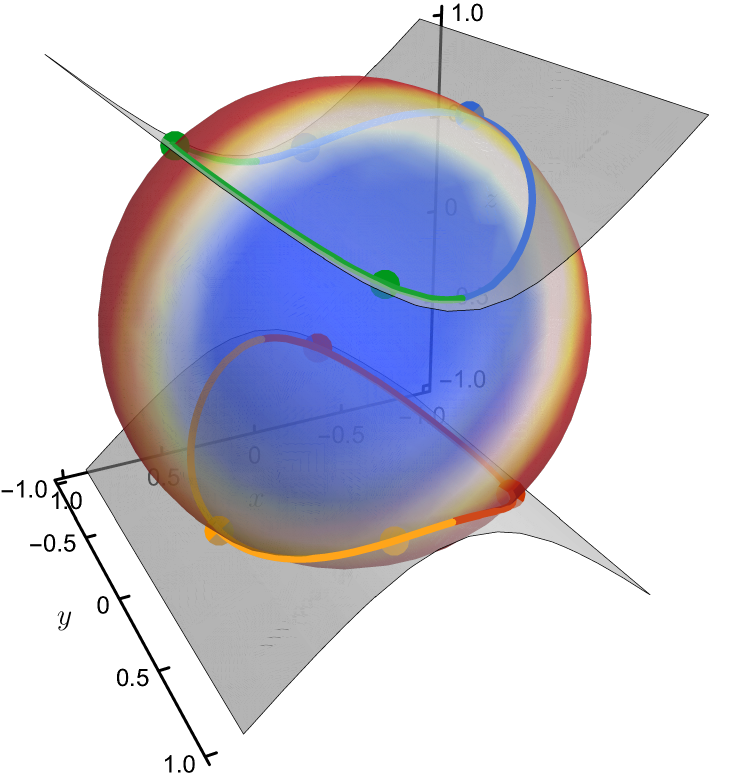}
	\caption{Illustration of the Levy constrained-search procedure for $t=1/2$, $U=1$, and $\rho=1/5$. 
	The value of $T+V_{ee}$ is mapped on the surface of the unit sphere that represents the normalized wave functions.
	The gray parabolas correspond to densities $\rho = z^2 - x^2$.
	The four branches of $f_{\pm\pm}$ [see Eq.~\eqref{eq:fpm}] are represented as contours and correspond to the intersections of these three-dimensional objects.
	The dots locate the stationary points on each of these contours.
	}
	\label{fig:LL}
\end{figure}

Substituting $x$ and $z$ in Eqs.~\eqref{eq:T} and \eqref{eq:Vee} thanks to the normalization condition and the reduced site occupation difference defined in Eqs.~\eqref{eq:normalization} and \eqref{eq:rho}, respectively, we obtain the four-branch function 
\begin{multline}
	\label{eq:fpm}
	f_{\pm\pm}(\rho,y) 
	= - 2 t y \qty(\pm\sqrt{1 - y^2 - \rho} \pm \sqrt{1 - y^2 + \rho}) 
	\\
	+ U \qty(1 - y^2)
\end{multline}
that one would minimize with respect to $y$ to obtain the exact ground-state functional. \cite{Schonhammer_1987,Carrascal_2015,Cohen_2016} Although one technically deals with \textit{functions} in the Hubbard dimer, we shall stick to the term \textit{functional} to emphasize the formal analogy between site-occupation function theory and DFT, as customarily done in the literature.\cite{Capelle_2013,Dimitrov_2016,Cohen_2016,Senjean_2017,Giarrusso_2022,Liebert_2023b}

Rather than only minimizing Eq.~\eqref{eq:fpm} for a given $\rho$, we seek \textit{all} stationary points \cite{Gorling_1999} of $f_{\pm\pm}(\rho,y)$ with respect to $y$, \ie,
\begin{equation}\label{eq:FnstatLL}
	F(\rho) =  \stat{y}\qty[f_{\pm\pm}(\rho,y)]
\end{equation}
The choice of the variable over which to optimize in Eq.~\eqref{eq:FnstatLL} is arbitrary and several choices are possible yielding various functions \cite{Carrascal_2015} other than $f_{\pm\pm}$, yet identical $F(\rho)$'s.
A similar procedure can be carried out via an ensemble formalism, \cite{Gross_1988a,Gross_1988b,Oliveira_1988} as shown by Fromager and coworkers. \cite{Deur_2017,Deur_2018,Deur_2019}

\alert{Because, taken as whole, $f_{\pm\pm}$ is symmetric with respect to a change in sign of $y$}, we restrict the discussion to the domain where $y \ge 0$, without loss of generality. As shown in Fig.~\ref{fig:fpm}, the branches $f_{++}$ and $f_{--}$ have one stationary point each for $y \ge 0$ (green square and red circle, respectively): the global minimum located at $y_0$ corresponds to the convex ground-state functional, $F_0(\rho) = f_{++}(\rho,y_0)$, while the global maximum at $y_2$ corresponds to the concave doubly-excited-state functional, \ie, $F_2(\rho) = f_{--}(\rho,y_2)$ (see Fig.~\ref{fig:F}). $F_0(\rho)$ and $F_2(\rho)$ merge at $\rho = 1$. The stationary points located at $-y_0$ and $-y_2$ are associated with opposite values of $\Delta v$.

For $\rho < \rho_\text{c}$, the branch $f_{+-}$ has two stationary points (yellow diamonds): a local minimum at $y_1^{\cap}$ and a local maximum at $y_1^{\cup}$ that yield a concave branch $F_1^{\cap}(\rho) = f_{+-}(\rho,y_1^{\cap})$ (yellow curve in Fig.~\ref{fig:F}) and a convex branch $F_1^{\cup}(\rho) = f_{+-}(\rho,y_1^{\cup})$ (blue curve in Fig.~\ref{fig:F}) for the singly-excited-state functional. As expected though, $F_1^{\cap}(\rho)$ and $F_1^{\cup}(\rho)$ lead to convex and concave energies, $E_1^{\cup}$ and $E_1^{\cap}$ (see Fig.~\ref{fig:energies}), respectively, preserving the property that the energy and the functional are conjugate functions. \cite{Helgaker_2022} \alert{Because the density-potential mapping associated with the first excited state is non-invertible (since, as seen in Fig.~\ref{fig:densities}, the same density $\rho$ can be produced by two $\Dv$ values)}, its ``functional'' is a partial (\ie, defined for a subdomain of $\rho$), multi-valued function constituted of one concave and one convex branch that correspond to two separate domains of the external potential. Again, the stationary points on $f_{-+}$ located at $-y_1^{\cap}$ and $-y_1^{\cup}$ (blue triangles) are associated with opposite values of $\Delta v$. At $\rho = \rho_\text{c}$, $y_1^{\cap}$ and $y_1^{\cup}$ merge and disappear for larger $\rho$ values. This critical value of the density decreases with respect to $U$ to reach zero at $U = 0$, and $\rho_\text{c} \to 1$ as $U\to\infty$. 

In accordance with Eq.~\eqref{eq:EulerDFT}, the derivative of $F_0(\rho)$ with respect to $\rho$ gives back $\Dv_0$ as a function of $\rho$, \ie, the inverse of $\rho_0 (\Delta v)$ plotted in Fig.~\ref{fig:densities}. Most notably, an analogous relation holds for the excited states. For the doubly-excited state, we simply have
\begin{equation}\label{eq:F2}
	\dv{F_2(\rho)}{\rho} = -\Dv_2 (\rho)
\end{equation}
In particular, for $\rho = 0$, we have $\Dv_2 = 0$, while $\Dv_2 \to \infty$ as $\rho \to 1$, similarly to $\Dv_0$ (except that $\Dv_0 \to -\infty$ as $\rho \to 1$).

For the first-excited state, which has a non-invertible density, $\rho_1(\Delta v)$ (see Fig.~\ref{fig:densities}), we still have 
\begin{subequations}
\begin{align}
	\label{eq:F1cup}
	&\dv{F_1^{\cup}(\rho)}{\rho} = - \Dv_1^{\cup}(\rho)
	\\
	\label{eq:F1cap}
	&\dv{F_1^{\cap}(\rho)}{\rho} = -\Dv_1^{\cap}(\rho)
\end{align}
\end{subequations}
where $\Dv_1^{\cup}(\rho)$ ranges from $-\Dv_c$ (at $\rho = \rho_c$) to $0^-$ (for $\rho \to 0^+$), yielding the inverse of the blue curve in Fig.~\ref{fig:densities}, and $\Dv_1^{\cap}(\rho)$ ranges from $-\infty$ (for $\rho \to 0^+$)  to $-\Dv_c$ (for $\rho = \rho_c$), yielding the inverse of the yellow curve in Fig.~\ref{fig:densities}.

The Levy constrained-search procedure is geometrically illustrated in Fig.~\ref{fig:LL}. The surface of the unit sphere corresponds to the normalized wave functions such that $x^2 + y^2 + z^2 = 1$, onto which we have mapped the value of $T + V_{ee}$ as a function of $x$, $y$, and $z$. The gray parabolas correspond to the (potentially unnormalized) wave functions yielding $\rho = z^2 - x^2$. Hence, the contours obtained by the intersection of these three-dimensional surfaces are the normalized wave functions yielding $\rho = z^2 - x^2$. On these contours, one is looking for the points where $f_{\pm\pm}$ is stationary. These are represented by the colored dots in Fig.~\ref{fig:LL} (see also Fig.~\ref{fig:fpm}).

\begin{figure}
	\includegraphics[width=\linewidth]{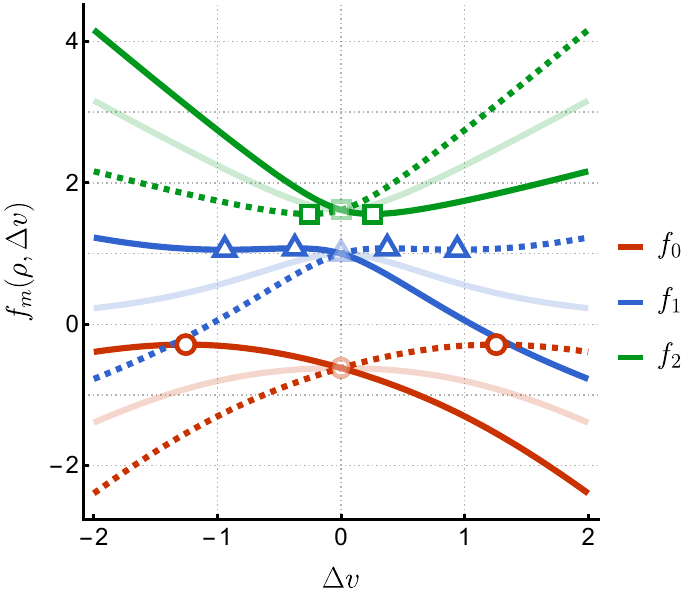}
	\caption{$f_m(\rho,\Dv)$ as a function of $\Dv$ for $t = 1/2$, $U = 1$, and $\rho=\pm1/2$: ground state ($m=0$), singly-excited state ($m=1$), and doubly-excited state ($m=2$). 
	The markers indicate the position of the stationary points. 
	The transparent curves correspond to $\rho=0$. 
	In this case, the linear term $-\Dv \rho$ in Eq.~\eqref{eq:Lvar} vanishes and the energy $E_m$ is recovered (see Fig.~\ref{fig:energies}). 
	For $\rho = 1/2$ (solid curves), the linear term shifts the maxima of $E_0$ and $E_1$ (red circle and blue triangle, respectively) towards $\Delta v<0$ and the minimum of $E_2$ (green square) towards $\Delta v>0$. 
	Moreover, a local minimum in $f_{1}$ (outermost blue triangle) appears. 
	The situation is exactly mirrored for $\rho = -1/2$ (dashed curves).}
	\label{fig:fn}
\end{figure}

The exact functionals represented in Fig.~\ref{fig:F} can also be obtained using the Lieb variational principle. To do so, let us define, for each singlet state, the function 
\begin{equation}
	f_{m}(\rho,\Dv) = E_m - \Dv \rho
\end{equation}
However, instead of maximizing the previous expression for a given $\rho$ as in Eq.~\eqref{eq:Lvar}, we seek its entire set of stationary points with respect to $\Dv$ for each $m$ value, \ie
\begin{equation}\label{eq:FnstatLdual}
	F_m(\rho) =\stat{\Dv}\qty[f_{m}(\rho,\Dv)]
\end{equation}

Figure \ref{fig:fn} shows $f_m$ as a function of $\Dv$ at $\rho = 0$ and $\pm 1/2$ for each state and the location of the corresponding stationary points. For $\rho=0$ (transparent curves), one recovers the energies $E_m$ plotted in Fig.~\ref{fig:energies}. The values of the functions $f_m$ at their stationary points (red circle, blue triangle, and green square at $\Dv = 0$) correspond to the initial values of $F_0$, $F_1^\cup$, and $F_2$ in Fig~\ref{fig:F}. For $\rho = 1/2$,  $f_0$ (solid red curve) and $f_2$ (solid green curve) have a single extremum: a maximum and a minimum yielding the ground- and second-excited-state functionals, $F_0(\rho)$ and $F_2(\rho)$, respectively, as depicted in Fig.~\ref{fig:F}. The blue curve $f_1$ exhibits a local maximum and minimum that correspond to the two branches of the multi-valued functional associated with the first-excited state, $F_1^{\cap}(\rho)$ and $F_1^{\cup}(\rho)$, respectively.

In practice, Lieb's formulation has a very neat geometric illustration in the Hubbard dimer. The total energies $E_m$ are ``tipped'' by the addition of the linear term $-\Dv  \rho$, which shifts their extrema: the maxima of $E_0$ and $E_1$ towards $\Delta v < 0$ and the minimum of $E_2$ towards $\Delta v >0$. Moreover, in the case of the first-excited state, the linear curve $- \Dv \rho$ shifts the energy in such a way that, as soon as $\rho>0$, a local minimum appears (outermost blue triangle) in $f_1$. This minimum and the maximum gradually get closer as $\rho$ increases, until they merge at $\rho = \rho_c$, $f_1$ becoming monotonic with no stationary points for $\rho > \rho_c$. The situation is exactly mirrored for $\rho = -1/2$ (dashed curves).

The present Letter reports the exact functional for the ground and (singlet) excited states of the asymmetric Hubbard dimer at half-filling. To the best of our knowledge, this is the first time that exact function(al)s corresponding to singlet (non-degenerate) excited states are computed. While the ground-state functional is well-known to be a convex function with respect to the site-occupation difference, the functional associated with the highest doubly-excited state is found to be concave. Additionally, and more importantly, we find that the ``functional'' for the first-excited state is a partial, multi-valued function of the density that is constructed from one concave and one convex branch associated with two separate domains of the external potential. Finally, Levy's constrained search and Lieb's convex formulation are found to be entirely consistent with one another, yielding the same exact functionals [Eqs.~\eqref{eq:FnstatLL} and \eqref{eq:FnstatLdual}] and, even more remarkably, the duality properties of the ground state appear to be \alert{shared} by the excited states of this model. These findings may provide insight into the challenges of constructing state-specific excited-state density functionals for general applications in electronic structure theory.

This project has received funding from the European Research Council (ERC) under the European Union's Horizon 2020 research and innovation programme (Grant agreement No.~863481).


\bibliography{HKvsLieb}

\end{document}